\pdfoutput=1

\documentclass[11pt]{article}

\usepackage[final]{ACL2023}

\usepackage{times}
\usepackage{latexsym}

\usepackage[T1]{fontenc}

\usepackage[utf8]{inputenc}

\usepackage{microtype}

\usepackage{inconsolata}
\usepackage{booktabs}
\usepackage{xspace}
\usepackage{amsmath}
\usepackage{amssymb}
\usepackage{mathtools}
\usepackage{tcolorbox}

\newcommand{\ours}{\texttt{HyDE}\xspace}
\newcommand{\InstructGPT}{\texttt{InstructGPT}\xspace}
\newcommand{\Contriever}{\texttt{Contriever}\xspace}
\newcommand{\mContriever}{\texttt{mContriever}\xspace}
\newcommand{\ContrieverFT}{\texttt{Contriever$^\text{FT}$}\xspace}
\newcommand{\mContrieverFT}{\texttt{mContriever$^\text{FT}$}\xspace}

\title{Precise Zero-Shot Dense Retrieval without Relevance Labels}

\author{Luyu Gao\thanks{~~~Equal contribution.}~~$^\dag$ \quad Xueguang Ma\footnotemark[1]~~$^\ddag$ \quad Jimmy Lin$^\ddag$ \quad Jamie Callan$^\dag$\\
$^\dag$Language Technologies Institute, Carnegie Mellon University\\
$^\ddag$David R. Cheriton School of Computer Science, University of Waterloo \\
\{luyug, callan\}@cs.cmu.edu,
\{x93ma, jimmylin\}@uwaterloo.ca
}

\begin{document}
\maketitle

\begin{abstract}
While dense retrieval has been shown effective and efficient across tasks and languages, it remains difficult to create effective fully zero-shot dense retrieval systems when no relevance label is available. 
In this paper, we recognize the difficulty of zero-shot learning and encoding relevance. Instead, we propose to pivot through \underline{Hy}pothetical \underline{D}ocument \underline{E}mbeddings~(\ours).
Given a query, \ours first zero-shot instructs an instruction-following language model~(e.g. \InstructGPT) to generate a \emph{hypothetical} document.
The document captures relevance patterns but is unreal and may contain false details. Then, an unsupervised contrastively learned encoder~(e.g. \Contriever) encodes the document into an embedding vector. This vector identifies a neighborhood in the corpus embedding space, where similar \emph{real} documents are retrieved based on vector similarity. This second step ground the generated document to the actual corpus, with the encoder's dense bottleneck filtering out the incorrect details.
Our experiments show that 
\ours significantly outperforms the state-of-the-art unsupervised dense retriever \Contriever and shows strong performance comparable to fine-tuned retrievers, across various tasks~(e.g. web search, QA, fact verification) and languages~(e.g. sw, ko, ja).\footnote{No models were trained or fine-tuned in making this pre-print. Our open source code is available at \url{https://github.com/texttron/hyde}.}
\end{abstract}
\section{Introduction}
Dense retrieval~\cite{lee-etal-2019-latent, karpukhin-etal-2020-dense}, the method of retrieving documents using semantic embedding similarities, has been shown successful across tasks like web search, question answering, and fact verification. A variety of methods such as negative mining~\cite{ance, qu-etal-2021-rocketqa}, distillation~\cite{qu-etal-2021-rocketqa, tct_colbert, tas-b} and task-specific pre-training~\cite{contriever, gao-callan-2021-condenser, lu-etal-2021-less, gao-callan-2022-unsupervised, Liu2022RetroMAEPR} have been proposed to improve the effectiveness of supervised dense retrieval models.

On the other hand, zero-shot dense retrieval still remains difficult. Many recent works consider the alternative transfer learning setup, where the dense retrievers are trained on a high-resource dataset and then evaluated on queries from new tasks.
The MS-MARCO collection~\cite{msmarco}, a massive judged dataset with a large number of judged query-document pairs, is arguably the most commonly used. As argued by \citet{contriever}, in practice, however, the existence of such a large dataset cannot always be assumed. Even MS-MARCO restricts commercial use and cannot be adopted in a variety of real-world search scenarios. 

\begin{figure*}[t]
  \includegraphics[width=\textwidth]{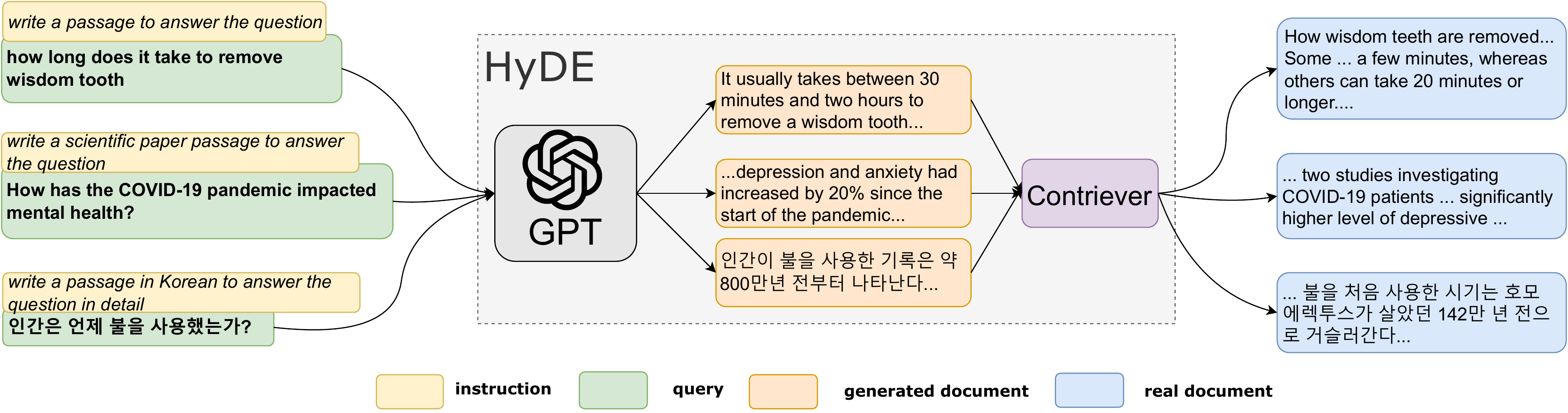}
  \caption{An illustration of the \ours model. Documents snippets are shown. \ours serves all types of queries without changing the underlying \texttt{GPT-3} and \Contriever/\mContriever models.}
  \label{fig:hyde}
\end{figure*}

In this paper, we aim to build effective fully zero-shot dense retrieval systems that require \textbf{no relevance} supervision, work out-of-box and generalize across tasks. As supervision is not available, we start by examining self-supervised representation learning methods. Modern deep learning enables two distinct learning algorithms. At the token level, generative large language models~(LLM) pre-trained on large corpus have demonstrated strong natural language understanding~(NLU) and generation~(NLG) capabilities~\cite{GPT-3, codex, Gopher, Chinchilla, LaMDA, palm}. At the document level, text (chunk) encoders pre-trained with contrastive objectives learn to encode document-document similarity into inner-product~\cite{contriever, gao-callan-2022-unsupervised}. On top of these, one extra insight into LLM is borrowed: the LLMs further trained to follow instructions can \emph{zero-shot} generalize to diverse unseen instructions~\cite{InstructGPT, T0, min-etal-2022-metaicl, FLAN}. \citet{InstructGPT} show that with a small amount of data, GPT-3~\cite{GPT-3} models can be aligned to human intent to follow instructions.

With these ingredients, we propose to pivot through \underline{Hy}pothetical \underline{D}ocument \underline{E}mbeddings~(\ours), and decompose dense retrieval into two tasks, a generative task performed by an instruction-following language model and a document-document similarity task performed by a contrastive encoder~(\autoref{fig:hyde}). First, we feed the query to the generative model and instruct it to "write a document that answers the question", i.e. a hypothetical document. We expect the generative process to capture "relevance" by giving an example; the generated document \textbf{is not} real, can contain factual errors but is like a relevant document. In the second step, we use an unsupervised contrastive encoder to encode this document into an embedding vector. Here, we expect the encoder's dense bottleneck to serve a lossy compressor, where the extra (hallucinated) details are filtered out from the embedding. We use this vector to search against the corpus embeddings. The most similar \emph{real} documents are retrieved and returned. The retrieval leverages document-document similarity encoded in the inner-product during contrastive training.
Note that, interestingly, with \ours factorization, the query-document similarity score is no longer explicitly modeled nor computed. Instead, the retrieval task is cast into two NLU and NLG tasks. 

\ours appears unsupervised. \textbf{No} model is trained in \ours: both the generative model and the contrastive encoder remain intact. Supervision signals were only involved in instruction learning of our backbone LLM. 

In our experiments, we show \ours using InstructGPT~\cite{InstructGPT} and Contriever~\cite{contriever} as backbone models significantly outperforms the previous state-of-the-art Contriever-only zero-shot no-relevance system on 11 queries sets, covering tasks like Web Search, Question Answering, Fact Verification and languages like Swahili, Korean, Japanese.
\section{Related Works}
\paragraph{Dense Retrieval}~\cite{lee-etal-2019-latent, karpukhin-etal-2020-dense} has been extensively studied after the emergence of pre-trained Transformer language models~\cite{devlin-etal-2019-bert}. Researchers studied the metric learning problems, such as training loss~\cite{karpukhin-etal-2020-dense} and negative sampling~\cite{ance, qu-etal-2021-rocketqa}, and also introduced distillation~\cite{qu-etal-2021-rocketqa, tct_colbert, tas-b}. Later works studied the second stage pre-training of language model specifically for retrieval~\cite{contriever, gao-callan-2021-condenser, lu-etal-2021-less, gao-callan-2022-unsupervised, Liu2022RetroMAEPR}.

The popularity of dense retrieval can be partially attributed to the rich and successful research in very efficient minimum inner product search~(MIPS) at very large~(billion) scales~\cite{JohnsonDJ17}.

\paragraph{Instructions-Following Language Models} Soon after the emergence of LLMs, several groups of researchers discover that LLMs trained on data consisting of instructions and their execution can zero-shot generalize to perform new tasks with new instructions~\cite{InstructGPT, T0, min-etal-2022-metaicl, FLAN}. This can be done by standard supervised sequence-to-sequence learning or more effectively with reinforcement learning~\cite{InstructGPT}.  

Concurrent to us, \citet{tart} studied ``Task-aware Retrieval with Instructions''. They \emph{fine-tuned dense encoders} that can also encode task-specific instruction prepended to query. In comparison, we use an unsupervised encoder and handle different tasks and their instruction with an instruction following generative LLM, as described above.

\paragraph{Zero-Shot Dense Retrieval} The tasks of zero-shot (dense) retrieval are arguably empirically defined by \citet{beir} for the neural retrieval community. Their BEIR benchmark consists of diverse retrieval tasks. The paper and many follow-up research generally consider the \texttt{Transfer Learning} setup where the dense retriever is first learned using a diverse and richly supervised corpus and query collection, namely MS-MARCO~\cite{beir, wang-etal-2022-gpl, yu2022cocodr}. 

However, as stated by \citet{contriever}, such a large collection can rarely be assumed. In this paper, therefore, we study the problem of building effective dense retrieval systems without relevance labels. Similar to \citet{contriever}, we also do not assume access to the test time corpora for training. This is a more realistic setup and prevents over-engineering on the test corpora.

By the definition in \citet{sachan2022improving}, our setup can be roughly considered as \textbf{``unsupervised''}. Strictly, as with \citet{sachan2022improving}, the only supervision resides in the LLM, in the processing of learning to follow instructions.

\paragraph{Generative Retrieval} Generative search is a new class of retrieval methods that use neural generative models as search indices~\cite{MetzlerTBN21, dsi, seal, arxiv.2204.13596}. These models use (constrained) decoding to generate document identifiers, such as id and sub-string, which map directly to \emph{real} documents. They have to go through special training procedures over relevance data; effective search may also need to use novel forms of search indices~\cite{seal, arxiv.2204.13596}. In comparison, our method uses the standard MIPS index and requires no training or training data. Our generative model produces an intermediate hypothetical document to be fed into a dense encoder, instead of a real document.
\section{Methodology}
In this section, we first formally define the problem of (zero-shot) dense retrieval. Then we will introduce how \ours is designed to solve it.
\subsection{Preliminaries}
Dense retrieval models similarity between query and document with inner product similarity. Given a query $q$ and document $d$, it uses two encoder function $\text{enc}_q$ and $\text{enc}_d$ to map them into $d$ dimension vectors $\mathbf{v_q}, \mathbf{v_d}$, whose inner product is used as similarity measurement.
\begin{equation}
    \text{sim}(\text{q}, \text{d}) 
    = \langle \text{enc}_q(\text{q}), \text{enc}_d(\text{d}) \rangle
    = \langle \mathbf{v_q}, \mathbf{v_d} \rangle
\label{eq:dense-retrieval}
\end{equation}
For zero-shot retrieval, we consider $L$ query sets $Q_1, Q_2, ..., Q_L$ and their corresponding search corpus, document sets $D_1, D_2, ..., D_L$. Denote the $j$-th query from $i$-th set query set $Q_i$ as $\text{q}_{ij}$. We need to fully define mapping \emph{functions} $\text{enc}_q$ and $\text{enc}_d$ without access to any query set $Q_i$, document set $D_i$, or any relevance judgment $r_{ij}$.

The difficulty of zero-shot dense retrieval lies precisely in \autoref{eq:dense-retrieval}: it requires learning of two embedding functions (for query and document respectively) into the \emph{same} embedding space where inner product captures \emph{relevance}. Without relevance judgments/scores to fit, learning becomes intractable.

\subsection{HyDE}
\ours circumvents the aforementioned learning problem by performing search in document-only embedding space that captures document-document similarity. This can be easily learned using unsupervised contrastive learning~\cite{contriever,gao-etal-2021-simcse,gao-callan-2022-unsupervised}. We set document encoder $\text{enc}_d$ directly as a contrastive encoder $\text{enc}_\text{con}$.
\begin{equation}
    f = \text{enc}_d = \text{enc}_\text{con}
\end{equation}
This function is also denoted as $f$ for simplicity. This unsupervised contrastive encoder will be shared by all incoming document corpus.
\begin{equation}
    \mathbf{v_d} = f(d) \quad \forall d \in D_1 \cup D_2 \cup ... \cup D_L
\end{equation}
To build the query vector, we consider in addition an instruction following LM, InstructLM. It takes a query $q$ and a textual instruction $\textsc{inst}$ and follows them to perform the task specified by $\textsc{inst}$. For simplicity, denote,
\begin{equation}
    g(q, \; \textsc{inst}) = \text{InstructLM} (q, \; \textsc{inst})
\end{equation}
Now we can use $g$ to map queries to "hypothetical" documents by sampling from $g$, setting $\textsc{inst}$ to be \texttt{``write a paragraph that answers the question''}. The generated document \emph{is not} real, can and is likely to be ungrounded factually~\cite{GPT-3, LaMDA}. We \emph{only} require it to capture relevance pattern. This is done by generating documents, i.e. providing examples. Critically, here we \textbf{offload} relevance modeling from representation learning model to an NLG model that generalizes significantly more easily, naturally, and effectively~\cite{GPT-3, InstructGPT}. Generating examples also replaces explicit modeling of relevance scores.
\newline
We can now encode the generated document using the document encoder $f$. Write,
\begin{equation}
    \mathbb{E}[\mathbf{v}_{q_{ij}}] = \mathbb{E}[f(g(q_{ij}, \textsc{inst}_i))]
    \label{eq:query-vec-exp}
\end{equation}
Formally, $g$ defines a probability distribution based on the chain rule. In this paper, we simply consider the expectation value, assuming the distribution of $\mathbf{v}_{q_{ij}}$ is uni-modal, i.e. the query is not ambiguous. The study of ambiguous queries and diversity is left to future work.
We estimate \autoref{eq:query-vec-exp} by sampling $N$ documents from $g$, $[\hat{d_1}, \hat{d_2}, ..., \hat{d_N}]$.
\begin{align}
    \hat{\mathbf{v}}_{q_{ij}}
    &= \frac{1}{N} \sum_{\hat{d_k} \sim g(q_{ij}, \textsc{inst}_i)} f(d_k) \\
    &= \frac{1}{N} \sum_{k=1}^N f (\hat{d_k})
\end{align}
We also consider the query as a possible hypothesis,
\begin{equation}
    \hat{\mathbf{v}}_{q_{ij}} = \frac{1}{N + 1} [\sum_{k=1}^N f (\hat{d_k}) + f(q_{ij})]
\end{equation}
Inner product is computed between $\hat{\mathbf{v}}_{q_{ij}}$ and the set of all document vectors $\{f(d) | d \in D_i\}$. The most similar documents are retrieved. Here the encoder function $f$ serves as a lossy compressor that outputs dense vectors, where the extra details are filtered and left out from the vector. It further grounds the hypothetical vector to the actual corpus and the real documents. The full \ours system is illustrated in \autoref{fig:hyde}.
\begin{table*}[h]
\centering
\small
\begin{tabular}{l|ccc|ccc}
\toprule
 & \multicolumn{3}{c|}{DL19}          & \multicolumn{3}{c}{DL20} \\
 & map  & ndcg@10 & recall@1k & map  & ndcg@10 & recall@1k \\
\midrule
\multicolumn{7}{l}{\textit{w/o relevance judgement}} \\
BM25 & 30.1 & 50.6 & 75.0 & 28.6 & 48.0 & 78.6 \\
Contriever & 24.0 & 44.5 & 74.6 & 24.0 & 42.1 & 75.4 \\
\ours & \textbf{41.8} & \textbf{61.3} & \textbf{88.0}	& \textbf{38.2}	& \textbf{57.9} & \textbf{84.4} \\
\midrule
\multicolumn{7}{l}{\textit{w/ relevance judgement}} \\
DPR  & 36.5 & 62.2 & 76.9 & 41.8 & \textbf{65.3} & 81.4 \\
ANCE & 37.1 & \textbf{64.5} & 75.5 & 40.8 & 64.6 & 77.6 \\
Contriever$^\text{FT}$ & 41.7 & 62.1 & 83.6 & \textbf{43.6} & 63.2 & \textbf{85.8} \\
\bottomrule
\end{tabular}
\caption{Results for web search on DL19/20. Best performing w/o relevance and overall system(s) are marked \textbf{bold}. DPR, ANCE and Contriever$^\text{FT}$ are in-domain \emph{supervised} models that are finetuned on MS MARCO training data.}
\label{tab:dl}
\end{table*}

\section{Experiments}
\subsection{Setup}
\paragraph{Implementation} We implement \ours using \InstructGPT, a GPT-3 model from the instruct series~(\texttt{text-davinci-003}; \citet{InstructGPT}) and \Contriever models~\cite{contriever}. We sample from \InstructGPT using the OpenAI playground default temperature of 0.7 for open-ended generations. We use the English-only \Contriever model for English retrieval tasks and multilingual \mContriever for non-English tasks. We conducted retrieval experiments with the Pyserini toolkit~\cite{Lin_etal_SIGIR2021_Pyserini}.

\paragraph{Datasets}
We consider web search query sets TREC DL19~\cite{dl19} and DL20~\cite{dl20}; they are based on the MS-MARCO dataset~\cite{msmarco}. We also use a diverse collection of 6 low-resource datasets from the BEIR dataset~\cite{beir}. For non-English retrieval, we consider Swahili, Korean, Japanese, and Bengali from the Mr.Tydi dataset~\cite{mrtydi}.

We use different instructions for each dataset. They share a similar structure but have different quantifiers to control the exact form of the generated hypothetical documents. These instructions can be found in \autoref{sec:app-inst}.
\paragraph{Compared Systems}
Contriever models, \Contriever and \mContriever, serve as our major baseline. They are trained using unsupervised contrastive learning.
\ours retrievers share the \emph{exact} same embedding spaces with them.
The only difference is how the query vector is built.
These comparisons allow us to easily examine the effect of \ours.
The classical heuristic-based lexical retriever BM25 is also included.

Several systems that involve fine-tuning on massive \emph{relevance} data are also included as references. We consider models fine-tuned on MS-MARCO and transferred, DPR and ANCE, from the BEIR paper.
For multilingual, we include the mDPR model from Mr.Tydi paper and MS-MARCO fine-tuned mBERT and XLM-R from the Contriever paper.
We also include the state-of-the-art transfer learning models: \Contriever and \mContriever fine-tuned on MS-MARCO, denoted \ContrieverFT and \mContrieverFT.
These models have run through the state-of-the-art retrieval model training pipeline that involves second-stage retrieval-specific pre-training~\cite{lee-etal-2019-latent} and a few rounds of fine-tuning~\cite{qu-etal-2021-rocketqa}; they should be considered empirical upper bounds.

\begin{table*}[]
\centering
\small
\begin{tabular}{l|cccccc}
\toprule
           & Scifact & Arguana & Trec-Covid & FiQA    & DBPedia & TREC-NEWS \\
\midrule
\multicolumn{7}{c}{\textit{nDCG@10}} \\
\multicolumn{7}{l}{\textit{w/o relevance judgement}} \\
BM25       & 67.9    & 39.7   & \textbf{59.5}     &  23.6 &  31.8   & 39.5  \\
Contriever & 64.9    & 37.9   & 27.3     &  24.5 &  29.2  &  34.8  \\
\ours      & \textbf{69.1}    & \textbf{46.6}   & 59.3     &  \textbf{27.3} &  \textbf{36.8}  &  \textbf{44.0}  \\
\midrule
\multicolumn{7}{l}{\textit{w/ relevance judgement}} \\
DPR        & 31.8    & 17.5    & 33.2      & 29.5    & 26.3    & 16.1 \\
ANCE       & 50.7    & 41.5    & \textbf{65.4}      & 30.0    & 28.1    & 38.2 \\
Contriever$^\text{FT}$ & 67.7 & 44.6 & 59.6 & \textbf{32.9} & \textbf{41.3} & 42.8 \\
\midrule
\multicolumn{7}{c}{\textit{Recall@100}} \\
\multicolumn{7}{l}{\textit{w/o relevance judgement}} \\
BM25       & 92.5   & 93.2   & \textbf{49.8}       & 54.0    & 46.8  & 44.7  \\
Contriever & 92.6   & 90.1   & 17.2      & 56.2    & 45.3  & 42.3  \\
\ours      & \textbf{96.4}   & \textbf{97.9}   & 41.4        & \textbf{62.1}    & \textbf{47.2}  & \textbf{50.9}  \\
\midrule
\multicolumn{7}{l}{\textit{w/ relevance judgement}} \\
DPR        & 72.7 & 75.1 & 21.2 & 34.2 & 34.9 &  21.5 \\
ANCE       & 81.6 & 93.7 & 45.7 & 58.1 & 31.9 &  39.8 \\
Contriever$^\text{FT}$ & 94.7 & 97.7 & 40.7 & \textbf{65.6} & \textbf{54.1} & 49.2 \\
\bottomrule
\end{tabular}
\caption{Low resource tasks from BEIR. Best performing w/o relevance and overall system(s) are marked \textbf{bold}.}
\label{tab:beir}
\end{table*}

\subsection{Web Search}
In \autoref{tab:dl}, we show retrieval results on TREC DL19 and TREC DL20. We see \ours bring sizable improvements to \Contriever across the board for both precision-oriented and recall metrics. While unsupervised \Contriever can underperform the classical BM25 approach, \ours outperforms BM25 by large margins.

\ours remains competitive even when compared to fine-tuned models. Note that TREC DL19/20 are search tasks defined on MS-MARCO and there, all the fine-tuned models are richly \emph{supervised}. On TREC DL19, \ours shows comparable map and ndcg@10 to \ContrieverFT and best recall@1k. On DL20, \ours gets around 10\% lower map and ndcg@10 than \ContrieverFT and similar recall@1k. The ANCE model shows better ndcg@10 numbers than \ours but lower recall, suggesting it may be biased to a subset of queries and/or relevant documents.

\subsection{Low Resource Retrieval}
In \autoref {tab:beir}, we show retrieval results on low-resource tasks from BEIR. Similar to web search, \ours again brings sizable improvements to \Contriever across the board in terms of both ndcg and recall. \ours is only outperformed by BM25 on one dataset, TREC-Covid but with a tiny 0.2 margin; in comparison, the underlying \Contriever underperforms by more than 50\%.

We also observe \ours demonstrates strong performance compared to fine-tuned models. \ours generally shows better performance than ANCE and DPR, even though the two are fine-tuned on MS-MARCO and ANCE also involves some sophisticated hard negative techniques. \ContrieverFT shows performance advantages on FiQA and DBPedia. These involve retrieval of financial posts or entities respectively. We believe the performance difference can be attributed to the under-specification of the instruction; more elaborative instructions may help.
\subsection{Multilingual Retrieval}
\begin{table}[]
\centering
\small
\resizebox{0.45\textwidth}{!}{
\begin{tabular}{l|cccc}
\toprule
            & Swahili & Korean & Japanese & Bengali\\
\midrule
\multicolumn{4}{l}{\textit{w/o relevance judgement}} \\
BM25        & 38.9  & 28.5    & 21.2 & \textbf{41.8}\\
mContriever & 38.3  & 22.3    & 19.5 & 35.3\\
\ours       & \textbf{41.7}  & \textbf{30.6}    & \textbf{30.7} & 41.3\\
\midrule
\multicolumn{4}{l}{\textit{w/ relevance judgement}} \\
mDPR  &  7.3  & 21.9 & 18.1 &  25.8 \\
mBERT &  37.4 & 28.1 & 27.1 &  35.1 \\
XLM-R &  35.1 & 32.2 & 24.8 &  41.7 \\
mContriever$^\text{FT}$ & \textbf{51.2} & \textbf{34.2} & \textbf{32.4} & \textbf{42.3}\\
\bottomrule
\end{tabular}}
\caption{MRR@100 on Mr.Tydi. Best performing w/o relevance and overall system(s) are marked \textbf{bold}.}
\label{tab:mrtydi}
\end{table}
Multilingual setup poses several additional challenges to \ours. The small-sized contrastive encoder gets saturated as the number of languages scales~\cite{conneau-etal-2020-unsupervised, contriever}. Meanwhile, our generative LLM faces an opposite issue: with languages of not as high resource as English or French, the high capacity LLM can get under-trained~\cite{Chinchilla}.

Nevertheless, in \autoref{tab:mrtydi}, we still find \ours able to improve the \mContriever model. It can outperform non-Contriever models fine-tuned on and transferred from MS-MARCO. On the other hand, we do observe some margins between \ours and fine-tuned \mContrieverFT. Since \ours and \mContrieverFT use similar contrastive encoders, we hypothesize this is because the non-English languages we considered are under-trained in both pre-training and instruction learning stages.
\section{Analysis}
The generative LLM and contrastive encoder make up the backbone of \ours. In this section, we study the effect of changing their realizations.  In particular, we consider smaller language models~(LM) and fine-tuned encoders. We conduct our studies on TREC DL19/20.

\begin{table}[h]
\centering
\small
\begin{tabular}{l|cc}
\toprule
Model                          & DL19 & DL20 \\
\midrule
Contriever                     & 44.5 & 42.1  \\
Contriever$^\text{FT}$         & 62.1 & 63.2  \\
\midrule
\ours \\
w/ Contriever                  &        &  \\
\hspace{10px} w/ Flan-T5 (11b)       & 48.9   & 52.9 \\
\hspace{10px} w/ Cohere (52b)        & 53.8   & 53.8 \\
\hspace{10px} w/ GPT (175b)          & \textbf{61.3}   & \textbf{57.9} \\
w/ Contriever$^{\text{FT}}$    &        &      \\
\hspace{10px} w/ Flan-T5 (11b)       & 60.2   & 62.1 \\
\hspace{10px} w/ Cohere (52b)        & 61.4   & 63.1 \\
\hspace{10px} w/ GPT (175b)          & \textbf{67.4}   & \textbf{63.5} \\
\bottomrule
\end{tabular}
\caption{NDCG@10 on TREC DL19/20. Effect of changing different instruction LMs and using fine-tuned encoder. Best w/o relevance and overall models are marked \textbf{bold}.}
\label{tab:model}
\end{table}

\subsection{Effect of Different Generative Models}
In \autoref{tab:model}, we show \ours using other instruction-following language models. In particular, we consider a 52-billion Cohere model~(\texttt{command-xlarge-20221108}) and a 11-billion FLAN model~(\texttt{FLAN-T5-xxl}; \citet{FLAN}).\footnote{Model sizes are from \url{https://crfm.stanford.edu/helm/v1.0/?models}.} Generally, we observe that all models bring improvement to the unsupervised \Contriever, with larger models bringing larger improvements. At the time when this paper is written, the Cohere model is still experimental without much detail disclosed. We can only tentatively hypothesize that training techniques may have also played some role in the performance difference.

\subsection{HyDE with Fine-tuned Encoder}
To begin with, \ours with fine-tuned encoder is \emph{not} the intended usage: \ours is more powerful and irreplaceable when few relevance labels are present. Here we are interested to find out if and how \ours embedding can affect fine-tuned encoders. In \autoref{tab:model}, we see that less powerful instruction LMs can negatively impact the overall performance of the fine-tuned retriever. (To remind our readers, \ContrieverFT is in-domain supervisedly fine-tuned for TREC DL19/20). The performance degradations remain small. On the other hand, we also observe the \InstructGPT model able to further bring up the performance, especially on DL19. This suggests that there may still exist certain factors not captured by the fine-tuned encoder but only by the generative model.
\section{Conclusion}
At the end of the paper, we encourage the readers to take a moment and reflect on the \ours model. Compare it to some of the other recently seen retrievers or re-ranker. These other models probably differ in their architecture, training method, and/or task, but probably all of them involve modeling relevance scores between a pair of query and document. Dense retrievers consider vector similarities while self-attentive re-rankers regression scores. In comparison, the concept of relevance in \ours is captured by an NLG model and the language generation process. We demonstrate in many cases, \ours can be as effective as dense retrievers that learn to model numerical relevance scores. So, is numerical relevance just a statistical artifact of language understanding? Will a weak retriever theoretically suffice as the NLU \& NLG models rapidly become stronger? Rushing to conclusions is not smart; more works need to be done to get answers. With this paper, we just want to raise these questions.

Concretely in this paper, we introduce a new paradigm of interactions between LLM and dense encoder/retriever. We demonstrate (part of) relevance modeling and instruction understanding can be delegated to the more powerful and flexible LLM. As a consequence, the need for relevance labels is removed. We are excited to see how this can be generalized further to more sophisticated tasks like multi-hop retrieval/QA and conversational search.

We argue \ours is also of practical use though not necessarily over the entire lifespan of a search system. At the very beginning of the life of the search system, serving queries using \ours offers performance comparable to a fine-tuned model, which no other relevance-free model can offer. As the search log grows, a supervised dense retriever can be gradually rolled out. As the dense retriever grows stronger, more queries will be routed to it, with only less common and emerging ones going to \ours backend.

\bibliography{anthology,custom}

\begin{thebibliography}{39}
\expandafter\ifx\csname natexlab\endcsname\relax\def\natexlab#1{#1}\fi

\bibitem[{Asai et~al.(2022)Asai, Schick, Lewis, Chen, Izacard, Riedel,
  Hajishirzi, and Yih}]{tart}
Akari Asai, Timo Schick, Patrick Lewis, Xilun Chen, Gautier Izacard, Sebastian
  Riedel, Hannaneh Hajishirzi, and Wen-tau Yih. 2022.
\newblock \href {https://doi.org/10.48550/ARXIV.2211.09260} {Task-aware
  retrieval with instructions}.

\bibitem[{Bajaj et~al.(2016)Bajaj, Campos, Craswell, Deng, Gao, Liu, Majumder,
  McNamara, Mitra, Nguyen, Rosenberg, Song, Stoica, Tiwary, and Wang}]{msmarco}
Payal Bajaj, Daniel Campos, Nick Craswell, Li~Deng, Jianfeng Gao, Xiaodong Liu,
  Rangan Majumder, Andrew McNamara, Bhaskar Mitra, Tri Nguyen, Mir Rosenberg,
  Xia Song, Alina Stoica, Saurabh Tiwary, and Tong Wang. 2016.
\newblock \href {https://doi.org/10.48550/ARXIV.1611.09268} {Ms marco: A human
  generated machine reading comprehension dataset}.

\bibitem[{Bevilacqua et~al.(2022)Bevilacqua, Ottaviano, Lewis, Yih, Riedel, and
  Petroni}]{seal}
Michele Bevilacqua, Giuseppe Ottaviano, Patrick Lewis, Wen{-}tau Yih, Sebastian
  Riedel, and Fabio Petroni. 2022.
\newblock \href {https://doi.org/10.48550/arXiv.2204.10628} {Autoregressive
  search engines: Generating substrings as document identifiers}.
\newblock \emph{CoRR}, abs/2204.10628.

\bibitem[{Brown et~al.(2020)Brown, Mann, Ryder, Subbiah, Kaplan, Dhariwal,
  Neelakantan, Shyam, Sastry, Askell, Agarwal, Herbert{-}Voss, Krueger,
  Henighan, Child, Ramesh, Ziegler, Wu, Winter, Hesse, Chen, Sigler, Litwin,
  Gray, Chess, Clark, Berner, McCandlish, Radford, Sutskever, and
  Amodei}]{GPT-3}
Tom~B. Brown, Benjamin Mann, Nick Ryder, Melanie Subbiah, Jared Kaplan,
  Prafulla Dhariwal, Arvind Neelakantan, Pranav Shyam, Girish Sastry, Amanda
  Askell, Sandhini Agarwal, Ariel Herbert{-}Voss, Gretchen Krueger, Tom
  Henighan, Rewon Child, Aditya Ramesh, Daniel~M. Ziegler, Jeffrey Wu, Clemens
  Winter, Christopher Hesse, Mark Chen, Eric Sigler, Mateusz Litwin, Scott
  Gray, Benjamin Chess, Jack Clark, Christopher Berner, Sam McCandlish, Alec
  Radford, Ilya Sutskever, and Dario Amodei. 2020.
\newblock \href
  {https://proceedings.neurips.cc/paper/2020/hash/1457c0d6bfcb4967418bfb8ac142f64a-Abstract.html}
  {Language models are few-shot learners}.
\newblock In \emph{Advances in Neural Information Processing Systems 33: Annual
  Conference on Neural Information Processing Systems 2020, NeurIPS 2020,
  December 6-12, 2020, virtual}.

\bibitem[{Chen et~al.(2021)Chen, Tworek, Jun, Yuan, Pinto, Kaplan, Edwards,
  Burda, Joseph, Brockman, Ray, Puri, Krueger, Petrov, Khlaaf, Sastry, Mishkin,
  Chan, Gray, Ryder, Pavlov, Power, Kaiser, Bavarian, Winter, Tillet, Such,
  Cummings, Plappert, Chantzis, Barnes, Herbert-Voss, Guss, Nichol, Paino,
  Tezak, Tang, Babuschkin, Balaji, Jain, Saunders, Hesse, Carr, Leike, Achiam,
  Misra, Morikawa, Radford, Knight, Brundage, Murati, Mayer, Welinder, McGrew,
  Amodei, McCandlish, Sutskever, and Zaremba}]{codex}
Mark Chen, Jerry Tworek, Heewoo Jun, Qiming Yuan, Henrique Ponde de~Oliveira
  Pinto, Jared Kaplan, Harri Edwards, Yuri Burda, Nicholas Joseph, Greg
  Brockman, Alex Ray, Raul Puri, Gretchen Krueger, Michael Petrov, Heidy
  Khlaaf, Girish Sastry, Pamela Mishkin, Brooke Chan, Scott Gray, Nick Ryder,
  Mikhail Pavlov, Alethea Power, Lukasz Kaiser, Mohammad Bavarian, Clemens
  Winter, Philippe Tillet, Felipe~Petroski Such, Dave Cummings, Matthias
  Plappert, Fotios Chantzis, Elizabeth Barnes, Ariel Herbert-Voss,
  William~Hebgen Guss, Alex Nichol, Alex Paino, Nikolas Tezak, Jie Tang, Igor
  Babuschkin, Suchir Balaji, Shantanu Jain, William Saunders, Christopher
  Hesse, Andrew~N. Carr, Jan Leike, Josh Achiam, Vedant Misra, Evan Morikawa,
  Alec Radford, Matthew Knight, Miles Brundage, Mira Murati, Katie Mayer, Peter
  Welinder, Bob McGrew, Dario Amodei, Sam McCandlish, Ilya Sutskever, and
  Wojciech Zaremba. 2021.
\newblock \href {https://doi.org/10.48550/ARXIV.2107.03374} {Evaluating large
  language models trained on code}.

\bibitem[{Chowdhery et~al.(2022)Chowdhery, Narang, Devlin, Bosma, Mishra,
  Roberts, Barham, Chung, Sutton, Gehrmann, Schuh, Shi, Tsvyashchenko, Maynez,
  Rao, Barnes, Tay, Shazeer, Prabhakaran, Reif, Du, Hutchinson, Pope, Bradbury,
  Austin, Isard, Gur-Ari, Yin, Duke, Levskaya, Ghemawat, Dev, Michalewski,
  Garcia, Misra, Robinson, Fedus, Zhou, Ippolito, Luan, Lim, Zoph, Spiridonov,
  Sepassi, Dohan, Agrawal, Omernick, Dai, Pillai, Pellat, Lewkowycz, Moreira,
  Child, Polozov, Lee, Zhou, Wang, Saeta, Diaz, Firat, Catasta, Wei,
  Meier-Hellstern, Eck, Dean, Petrov, and Fiedel}]{palm}
Aakanksha Chowdhery, Sharan Narang, Jacob Devlin, Maarten Bosma, Gaurav Mishra,
  Adam Roberts, Paul Barham, Hyung~Won Chung, Charles Sutton, Sebastian
  Gehrmann, Parker Schuh, Kensen Shi, Sasha Tsvyashchenko, Joshua Maynez,
  Abhishek Rao, Parker Barnes, Yi~Tay, Noam Shazeer, Vinodkumar Prabhakaran,
  Emily Reif, Nan Du, Ben Hutchinson, Reiner Pope, James Bradbury, Jacob
  Austin, Michael Isard, Guy Gur-Ari, Pengcheng Yin, Toju Duke, Anselm
  Levskaya, Sanjay Ghemawat, Sunipa Dev, Henryk Michalewski, Xavier Garcia,
  Vedant Misra, Kevin Robinson, Liam Fedus, Denny Zhou, Daphne Ippolito, David
  Luan, Hyeontaek Lim, Barret Zoph, Alexander Spiridonov, Ryan Sepassi, David
  Dohan, Shivani Agrawal, Mark Omernick, Andrew~M. Dai,
  Thanumalayan~Sankaranarayana Pillai, Marie Pellat, Aitor Lewkowycz, Erica
  Moreira, Rewon Child, Oleksandr Polozov, Katherine Lee, Zongwei Zhou, Xuezhi
  Wang, Brennan Saeta, Mark Diaz, Orhan Firat, Michele Catasta, Jason Wei,
  Kathy Meier-Hellstern, Douglas Eck, Jeff Dean, Slav Petrov, and Noah Fiedel.
  2022.
\newblock \href {https://doi.org/10.48550/ARXIV.2204.02311} {Palm: Scaling
  language modeling with pathways}.

\bibitem[{Conneau et~al.(2020)Conneau, Khandelwal, Goyal, Chaudhary, Wenzek,
  Guzm{\'a}n, Grave, Ott, Zettlemoyer, and
  Stoyanov}]{conneau-etal-2020-unsupervised}
Alexis Conneau, Kartikay Khandelwal, Naman Goyal, Vishrav Chaudhary, Guillaume
  Wenzek, Francisco Guzm{\'a}n, Edouard Grave, Myle Ott, Luke Zettlemoyer, and
  Veselin Stoyanov. 2020.
\newblock \href {https://doi.org/10.18653/v1/2020.acl-main.747} {Unsupervised
  cross-lingual representation learning at scale}.
\newblock In \emph{Proceedings of the 58th Annual Meeting of the Association
  for Computational Linguistics}, pages 8440--8451, Online. Association for
  Computational Linguistics.

\bibitem[{Craswell et~al.(2020{\natexlab{a}})Craswell, Mitra, Yilmaz, Campos,
  and Voorhees}]{dl19}
Nick Craswell, Bhaskar Mitra, Emine Yilmaz, Daniel Campos, and Ellen~M.
  Voorhees. 2020{\natexlab{a}}.
\newblock \href {https://doi.org/10.48550/ARXIV.2003.07820} {Overview of the
  trec 2019 deep learning track}.

\bibitem[{Craswell et~al.(2020{\natexlab{b}})Craswell, Mitra, Yilmaz, Campos,
  and Voorhees}]{dl20}
Nick Craswell, Bhaskar Mitra, Emine Yilmaz, Daniel~Fernando Campos, and
  Ellen~M. Voorhees. 2020{\natexlab{b}}.
\newblock Overview of the trec 2020 deep learning track.
\newblock \emph{ArXiv}, abs/2003.07820.

\bibitem[{Devlin et~al.(2019)Devlin, Chang, Lee, and
  Toutanova}]{devlin-etal-2019-bert}
Jacob Devlin, Ming-Wei Chang, Kenton Lee, and Kristina Toutanova. 2019.
\newblock \href {https://doi.org/10.18653/v1/N19-1423} {{BERT}: Pre-training of
  deep bidirectional transformers for language understanding}.
\newblock In \emph{Proceedings of the 2019 Conference of the North {A}merican
  Chapter of the Association for Computational Linguistics: Human Language
  Technologies, Volume 1 (Long and Short Papers)}, pages 4171--4186,
  Minneapolis, Minnesota. Association for Computational Linguistics.

\bibitem[{Gao and Callan(2021)}]{gao-callan-2021-condenser}
Luyu Gao and Jamie Callan. 2021.
\newblock \href {https://doi.org/10.18653/v1/2021.emnlp-main.75} {Condenser: a
  pre-training architecture for dense retrieval}.
\newblock In \emph{Proceedings of the 2021 Conference on Empirical Methods in
  Natural Language Processing}, pages 981--993, Online and Punta Cana,
  Dominican Republic. Association for Computational Linguistics.

\bibitem[{Gao and Callan(2022)}]{gao-callan-2022-unsupervised}
Luyu Gao and Jamie Callan. 2022.
\newblock \href {https://doi.org/10.18653/v1/2022.acl-long.203} {Unsupervised
  corpus aware language model pre-training for dense passage retrieval}.
\newblock In \emph{Proceedings of the 60th Annual Meeting of the Association
  for Computational Linguistics (Volume 1: Long Papers)}, pages 2843--2853,
  Dublin, Ireland. Association for Computational Linguistics.

\bibitem[{Gao et~al.(2021)Gao, Yao, and Chen}]{gao-etal-2021-simcse}
Tianyu Gao, Xingcheng Yao, and Danqi Chen. 2021.
\newblock \href {https://doi.org/10.18653/v1/2021.emnlp-main.552} {{S}im{CSE}:
  Simple contrastive learning of sentence embeddings}.
\newblock In \emph{Proceedings of the 2021 Conference on Empirical Methods in
  Natural Language Processing}, pages 6894--6910, Online and Punta Cana,
  Dominican Republic. Association for Computational Linguistics.

\bibitem[{Hoffmann et~al.(2022)Hoffmann, Borgeaud, Mensch, Buchatskaya, Cai,
  Rutherford, Casas, Hendricks, Welbl, Clark, Hennigan, Noland, Millican,
  Driessche, Damoc, Guy, Osindero, Simonyan, Elsen, Rae, Vinyals, and
  Sifre}]{Chinchilla}
Jordan Hoffmann, Sebastian Borgeaud, Arthur Mensch, Elena Buchatskaya, Trevor
  Cai, Eliza Rutherford, Diego de~Las Casas, Lisa~Anne Hendricks, Johannes
  Welbl, Aidan Clark, Tom Hennigan, Eric Noland, Katie Millican, George van~den
  Driessche, Bogdan Damoc, Aurelia Guy, Simon Osindero, Karen Simonyan, Erich
  Elsen, Jack~W. Rae, Oriol Vinyals, and Laurent Sifre. 2022.
\newblock \href {https://doi.org/10.48550/ARXIV.2203.15556} {Training
  compute-optimal large language models}.

\bibitem[{Hofst\"{a}tter et~al.(2021)Hofst\"{a}tter, Lin, Yang, Lin, and
  Hanbury}]{tas-b}
Sebastian Hofst\"{a}tter, Sheng-Chieh Lin, Jheng-Hong Yang, Jimmy Lin, and
  Allan Hanbury. 2021.
\newblock \href {https://doi.org/10.1145/3404835.3462891} {Efficiently teaching
  an effective dense retriever with balanced topic aware sampling}.
\newblock In \emph{Proceedings of the 44th International ACM SIGIR Conference
  on Research and Development in Information Retrieval}, SIGIR '21, page
  113–122, New York, NY, USA. Association for Computing Machinery.

\bibitem[{Izacard et~al.(2021)Izacard, Caron, Hosseini, Riedel, Bojanowski,
  Joulin, and Grave}]{contriever}
Gautier Izacard, Mathilde Caron, Lucas Hosseini, Sebastian Riedel, Piotr
  Bojanowski, Armand Joulin, and Edouard Grave. 2021.
\newblock \href {http://arxiv.org/abs/2112.09118} {Towards unsupervised dense
  information retrieval with contrastive learning}.
\newblock \emph{CoRR}, abs/2112.09118.

\bibitem[{Johnson et~al.(2017)Johnson, Douze, and J{\'{e}}gou}]{JohnsonDJ17}
Jeff Johnson, Matthijs Douze, and Herv{\'{e}} J{\'{e}}gou. 2017.
\newblock \href {http://arxiv.org/abs/1702.08734} {Billion-scale similarity
  search with gpus}.
\newblock \emph{CoRR}, abs/1702.08734.

\bibitem[{Karpukhin et~al.(2020)Karpukhin, Oguz, Min, Lewis, Wu, Edunov, Chen,
  and Yih}]{karpukhin-etal-2020-dense}
Vladimir Karpukhin, Barlas Oguz, Sewon Min, Patrick Lewis, Ledell Wu, Sergey
  Edunov, Danqi Chen, and Wen-tau Yih. 2020.
\newblock \href {https://doi.org/10.18653/v1/2020.emnlp-main.550} {Dense
  passage retrieval for open-domain question answering}.
\newblock In \emph{Proceedings of the 2020 Conference on Empirical Methods in
  Natural Language Processing (EMNLP)}, pages 6769--6781, Online. Association
  for Computational Linguistics.

\bibitem[{Lee et~al.(2022)Lee, Yang, Oh, and Seo}]{arxiv.2204.13596}
Hyunji Lee, Sohee Yang, Hanseok Oh, and Minjoon Seo. 2022.
\newblock \href {https://doi.org/10.48550/ARXIV.2204.13596} {Generative
  multi-hop retrieval}.

\bibitem[{Lee et~al.(2019)Lee, Chang, and Toutanova}]{lee-etal-2019-latent}
Kenton Lee, Ming-Wei Chang, and Kristina Toutanova. 2019.
\newblock \href {https://doi.org/10.18653/v1/P19-1612} {Latent retrieval for
  weakly supervised open domain question answering}.
\newblock In \emph{Proceedings of the 57th Annual Meeting of the Association
  for Computational Linguistics}, pages 6086--6096, Florence, Italy.
  Association for Computational Linguistics.

\bibitem[{Lin et~al.(2021{\natexlab{a}})Lin, Ma, Lin, Yang, Pradeep, and
  Nogueira}]{Lin_etal_SIGIR2021_Pyserini}
Jimmy Lin, Xueguang Ma, Sheng-Chieh Lin, Jheng-Hong Yang, Ronak Pradeep, and
  Rodrigo Nogueira. 2021{\natexlab{a}}.
\newblock {Pyserini}: A {Python} toolkit for reproducible information retrieval
  research with sparse and dense representations.
\newblock In \emph{Proceedings of the 44th Annual International ACM SIGIR
  Conference on Research and Development in Information Retrieval (SIGIR
  2021)}, pages 2356--2362.

\bibitem[{Lin et~al.(2021{\natexlab{b}})Lin, Yang, and Lin}]{tct_colbert}
Sheng-Chieh Lin, Jheng-Hong Yang, and Jimmy Lin. 2021{\natexlab{b}}.
\newblock \href {https://doi.org/10.18653/v1/2021.repl4nlp-1.17} {In-batch
  negatives for knowledge distillation with tightly-coupled teachers for dense
  retrieval}.
\newblock In \emph{Proceedings of the 6th Workshop on Representation Learning
  for NLP (RepL4NLP-2021)}, pages 163--173, Online. Association for
  Computational Linguistics.

\bibitem[{Liu and Shao(2022)}]{Liu2022RetroMAEPR}
Zheng Liu and Yingxia Shao. 2022.
\newblock Retromae: Pre-training retrieval-oriented transformers via masked
  auto-encoder.
\newblock \emph{ArXiv}, abs/2205.12035.

\bibitem[{Lu et~al.(2021)Lu, He, Xiong, Ke, Malik, Dou, Bennett, Liu, and
  Overwijk}]{lu-etal-2021-less}
Shuqi Lu, Di~He, Chenyan Xiong, Guolin Ke, Waleed Malik, Zhicheng Dou, Paul
  Bennett, Tie-Yan Liu, and Arnold Overwijk. 2021.
\newblock \href {https://doi.org/10.18653/v1/2021.emnlp-main.220} {Less is
  more: Pretrain a strong {S}iamese encoder for dense text retrieval using a
  weak decoder}.
\newblock In \emph{Proceedings of the 2021 Conference on Empirical Methods in
  Natural Language Processing}, pages 2780--2791, Online and Punta Cana,
  Dominican Republic. Association for Computational Linguistics.

\bibitem[{Metzler et~al.(2021)Metzler, Tay, Bahri, and Najork}]{MetzlerTBN21}
Donald Metzler, Yi~Tay, Dara Bahri, and Marc Najork. 2021.
\newblock \href {https://doi.org/10.1145/3476415.3476428} {Rethinking search:
  making domain experts out of dilettantes}.
\newblock \emph{{SIGIR} Forum}, 55(1):13:1--13:27.

\bibitem[{Min et~al.(2022)Min, Lewis, Zettlemoyer, and
  Hajishirzi}]{min-etal-2022-metaicl}
Sewon Min, Mike Lewis, Luke Zettlemoyer, and Hannaneh Hajishirzi. 2022.
\newblock \href {https://doi.org/10.18653/v1/2022.naacl-main.201} {{M}eta{ICL}:
  Learning to learn in context}.
\newblock In \emph{Proceedings of the 2022 Conference of the North American
  Chapter of the Association for Computational Linguistics: Human Language
  Technologies}, pages 2791--2809, Seattle, United States. Association for
  Computational Linguistics.

\bibitem[{Ouyang et~al.(2022)Ouyang, Wu, Jiang, Almeida, Wainwright, Mishkin,
  Zhang, Agarwal, Slama, Ray, Schulman, Hilton, Kelton, Miller, Simens, Askell,
  Welinder, Christiano, Leike, and Lowe}]{InstructGPT}
Long Ouyang, Jeff Wu, Xu~Jiang, Diogo Almeida, Carroll~L. Wainwright, Pamela
  Mishkin, Chong Zhang, Sandhini Agarwal, Katarina Slama, Alex Ray, John
  Schulman, Jacob Hilton, Fraser Kelton, Luke Miller, Maddie Simens, Amanda
  Askell, Peter Welinder, Paul Christiano, Jan Leike, and Ryan Lowe. 2022.
\newblock \href {https://doi.org/10.48550/ARXIV.2203.02155} {Training language
  models to follow instructions with human feedback}.

\bibitem[{Qu et~al.(2021)Qu, Ding, Liu, Liu, Ren, Zhao, Dong, Wu, and
  Wang}]{qu-etal-2021-rocketqa}
Yingqi Qu, Yuchen Ding, Jing Liu, Kai Liu, Ruiyang Ren, Wayne~Xin Zhao, Daxiang
  Dong, Hua Wu, and Haifeng Wang. 2021.
\newblock \href {https://doi.org/10.18653/v1/2021.naacl-main.466}
  {{R}ocket{QA}: An optimized training approach to dense passage retrieval for
  open-domain question answering}.
\newblock In \emph{Proceedings of the 2021 Conference of the North American
  Chapter of the Association for Computational Linguistics: Human Language
  Technologies}, pages 5835--5847, Online. Association for Computational
  Linguistics.

\bibitem[{Rae et~al.(2021)Rae, Borgeaud, Cai, Millican, Hoffmann, Song,
  Aslanides, Henderson, Ring, Young, Rutherford, Hennigan, Menick, Cassirer,
  Powell, Driessche, Hendricks, Rauh, Huang, Glaese, Welbl, Dathathri, Huang,
  Uesato, Mellor, Higgins, Creswell, McAleese, Wu, Elsen, Jayakumar,
  Buchatskaya, Budden, Sutherland, Simonyan, Paganini, Sifre, Martens, Li,
  Kuncoro, Nematzadeh, Gribovskaya, Donato, Lazaridou, Mensch, Lespiau,
  Tsimpoukelli, Grigorev, Fritz, Sottiaux, Pajarskas, Pohlen, Gong, Toyama,
  d'Autume, Li, Terzi, Mikulik, Babuschkin, Clark, Casas, Guy, Jones, Bradbury,
  Johnson, Hechtman, Weidinger, Gabriel, Isaac, Lockhart, Osindero, Rimell,
  Dyer, Vinyals, Ayoub, Stanway, Bennett, Hassabis, Kavukcuoglu, and
  Irving}]{Gopher}
Jack~W. Rae, Sebastian Borgeaud, Trevor Cai, Katie Millican, Jordan Hoffmann,
  Francis Song, John Aslanides, Sarah Henderson, Roman Ring, Susannah Young,
  Eliza Rutherford, Tom Hennigan, Jacob Menick, Albin Cassirer, Richard Powell,
  George van~den Driessche, Lisa~Anne Hendricks, Maribeth Rauh, Po-Sen Huang,
  Amelia Glaese, Johannes Welbl, Sumanth Dathathri, Saffron Huang, Jonathan
  Uesato, John Mellor, Irina Higgins, Antonia Creswell, Nat McAleese, Amy Wu,
  Erich Elsen, Siddhant Jayakumar, Elena Buchatskaya, David Budden, Esme
  Sutherland, Karen Simonyan, Michela Paganini, Laurent Sifre, Lena Martens,
  Xiang~Lorraine Li, Adhiguna Kuncoro, Aida Nematzadeh, Elena Gribovskaya,
  Domenic Donato, Angeliki Lazaridou, Arthur Mensch, Jean-Baptiste Lespiau,
  Maria Tsimpoukelli, Nikolai Grigorev, Doug Fritz, Thibault Sottiaux, Mantas
  Pajarskas, Toby Pohlen, Zhitao Gong, Daniel Toyama, Cyprien de~Masson
  d'Autume, Yujia Li, Tayfun Terzi, Vladimir Mikulik, Igor Babuschkin, Aidan
  Clark, Diego de~Las Casas, Aurelia Guy, Chris Jones, James Bradbury, Matthew
  Johnson, Blake Hechtman, Laura Weidinger, Iason Gabriel, William Isaac,
  Ed~Lockhart, Simon Osindero, Laura Rimell, Chris Dyer, Oriol Vinyals, Kareem
  Ayoub, Jeff Stanway, Lorrayne Bennett, Demis Hassabis, Koray Kavukcuoglu, and
  Geoffrey Irving. 2021.
\newblock \href {https://doi.org/10.48550/ARXIV.2112.11446} {Scaling language
  models: Methods, analysis \& insights from training gopher}.

\bibitem[{Sachan et~al.(2022)Sachan, Lewis, Joshi, Aghajanyan, Yih, Pineau, and
  Zettlemoyer}]{sachan2022improving}
Devendra~Singh Sachan, Mike Lewis, Mandar Joshi, Armen Aghajanyan, Wen-tau Yih,
  Joelle Pineau, and Luke Zettlemoyer. 2022.
\newblock \href {https://arxiv.org/abs/2204.07496} {Improving passage retrieval
  with zero-shot question generation}.

\bibitem[{Sanh et~al.(2022)Sanh, Webson, Raffel, Bach, Sutawika, Alyafeai,
  Chaffin, Stiegler, Raja, Dey, Bari, Xu, Thakker, Sharma, Szczechla, Kim,
  Chhablani, Nayak, Datta, Chang, Jiang, Wang, Manica, Shen, Yong, Pandey,
  Bawden, Wang, Neeraj, Rozen, Sharma, Santilli, F{\'{e}}vry, Fries, Teehan,
  Scao, Biderman, Gao, Wolf, and Rush}]{T0}
Victor Sanh, Albert Webson, Colin Raffel, Stephen Bach, Lintang Sutawika, Zaid
  Alyafeai, Antoine Chaffin, Arnaud Stiegler, Arun Raja, Manan Dey, M~Saiful
  Bari, Canwen Xu, Urmish Thakker, Shanya~Sharma Sharma, Eliza Szczechla,
  Taewoon Kim, Gunjan Chhablani, Nihal~V. Nayak, Debajyoti Datta, Jonathan
  Chang, Mike~Tian{-}Jian Jiang, Han Wang, Matteo Manica, Sheng Shen, Zheng~Xin
  Yong, Harshit Pandey, Rachel Bawden, Thomas Wang, Trishala Neeraj, Jos Rozen,
  Abheesht Sharma, Andrea Santilli, Thibault F{\'{e}}vry, Jason~Alan Fries,
  Ryan Teehan, Teven~Le Scao, Stella Biderman, Leo Gao, Thomas Wolf, and
  Alexander~M. Rush. 2022.
\newblock \href {https://openreview.net/forum?id=9Vrb9D0WI4} {Multitask
  prompted training enables zero-shot task generalization}.
\newblock In \emph{The Tenth International Conference on Learning
  Representations, {ICLR} 2022, Virtual Event, April 25-29, 2022}.
  OpenReview.net.

\bibitem[{Tay et~al.(2022)Tay, Tran, Dehghani, Ni, Bahri, Mehta, Qin, Hui,
  Zhao, Gupta, Schuster, Cohen, and Metzler}]{dsi}
Yi~Tay, Vinh~Q. Tran, Mostafa Dehghani, Jianmo Ni, Dara Bahri, Harsh Mehta,
  Zhen Qin, Kai Hui, Zhe Zhao, Jai~Prakash Gupta, Tal Schuster, William~W.
  Cohen, and Donald Metzler. 2022.
\newblock \href {http://arxiv.org/abs/2202.06991} {Transformer memory as a
  differentiable search index}.
\newblock \emph{CoRR}, abs/2202.06991.

\bibitem[{Thakur et~al.(2021)Thakur, Reimers, R{\"{u}}ckl{\'{e}}, Srivastava,
  and Gurevych}]{beir}
Nandan Thakur, Nils Reimers, Andreas R{\"{u}}ckl{\'{e}}, Abhishek Srivastava,
  and Iryna Gurevych. 2021.
\newblock \href {http://arxiv.org/abs/2104.08663} {{BEIR:} {A} heterogenous
  benchmark for zero-shot evaluation of information retrieval models}.
\newblock \emph{CoRR}, abs/2104.08663.

\bibitem[{Thoppilan et~al.(2022)Thoppilan, Freitas, Hall, Shazeer,
  Kulshreshtha, Cheng, Jin, Bos, Baker, Du, Li, Lee, Zheng, Ghafouri, Menegali,
  Huang, Krikun, Lepikhin, Qin, Chen, Xu, Chen, Roberts, Bosma, Zhou, Chang,
  Krivokon, Rusch, Pickett, Meier{-}Hellstern, Morris, Doshi, Santos, Duke,
  Soraker, Zevenbergen, Prabhakaran, Diaz, Hutchinson, Olson, Molina,
  Hoffman{-}John, Lee, Aroyo, Rajakumar, Butryna, Lamm, Kuzmina, Fenton, Cohen,
  Bernstein, Kurzweil, Aguera{-}Arcas, Cui, Croak, Chi, and Le}]{LaMDA}
Romal Thoppilan, Daniel~De Freitas, Jamie Hall, Noam Shazeer, Apoorv
  Kulshreshtha, Heng{-}Tze Cheng, Alicia Jin, Taylor Bos, Leslie Baker, Yu~Du,
  YaGuang Li, Hongrae Lee, Huaixiu~Steven Zheng, Amin Ghafouri, Marcelo
  Menegali, Yanping Huang, Maxim Krikun, Dmitry Lepikhin, James Qin, Dehao
  Chen, Yuanzhong Xu, Zhifeng Chen, Adam Roberts, Maarten Bosma, Yanqi Zhou,
  Chung{-}Ching Chang, Igor Krivokon, Will Rusch, Marc Pickett, Kathleen~S.
  Meier{-}Hellstern, Meredith~Ringel Morris, Tulsee Doshi, Renelito~Delos
  Santos, Toju Duke, Johnny Soraker, Ben Zevenbergen, Vinodkumar Prabhakaran,
  Mark Diaz, Ben Hutchinson, Kristen Olson, Alejandra Molina, Erin
  Hoffman{-}John, Josh Lee, Lora Aroyo, Ravi Rajakumar, Alena Butryna, Matthew
  Lamm, Viktoriya Kuzmina, Joe Fenton, Aaron Cohen, Rachel Bernstein, Ray
  Kurzweil, Blaise Aguera{-}Arcas, Claire Cui, Marian Croak, Ed~H. Chi, and
  Quoc Le. 2022.
\newblock \href {http://arxiv.org/abs/2201.08239} {Lamda: Language models for
  dialog applications}.
\newblock \emph{CoRR}, abs/2201.08239.

\bibitem[{Wang et~al.(2022)Wang, Thakur, Reimers, and
  Gurevych}]{wang-etal-2022-gpl}
Kexin Wang, Nandan Thakur, Nils Reimers, and Iryna Gurevych. 2022.
\newblock \href {https://doi.org/10.18653/v1/2022.naacl-main.168} {{GPL}:
  Generative pseudo labeling for unsupervised domain adaptation of dense
  retrieval}.
\newblock In \emph{Proceedings of the 2022 Conference of the North American
  Chapter of the Association for Computational Linguistics: Human Language
  Technologies}, pages 2345--2360, Seattle, United States. Association for
  Computational Linguistics.

\bibitem[{Wei et~al.(2022)Wei, Bosma, Zhao, Guu, Yu, Lester, Du, Dai, and
  Le}]{FLAN}
Jason Wei, Maarten Bosma, Vincent~Y. Zhao, Kelvin Guu, Adams~Wei Yu, Brian
  Lester, Nan Du, Andrew~M. Dai, and Quoc~V. Le. 2022.
\newblock \href {https://openreview.net/forum?id=gEZrGCozdqR} {Finetuned
  language models are zero-shot learners}.
\newblock In \emph{The Tenth International Conference on Learning
  Representations, {ICLR} 2022, Virtual Event, April 25-29, 2022}.
  OpenReview.net.

\bibitem[{Xiong et~al.(2021)Xiong, Xiong, Li, Tang, Liu, Bennett, Ahmed, and
  Overwijk}]{ance}
Lee Xiong, Chenyan Xiong, Ye~Li, Kwok{-}Fung Tang, Jialin Liu, Paul~N. Bennett,
  Junaid Ahmed, and Arnold Overwijk. 2021.
\newblock \href {https://openreview.net/forum?id=zeFrfgyZln} {Approximate
  nearest neighbor negative contrastive learning for dense text retrieval}.
\newblock In \emph{9th International Conference on Learning Representations,
  {ICLR} 2021, Virtual Event, Austria, May 3-7, 2021}. OpenReview.net.

\bibitem[{Yu et~al.(2022)Yu, Xiong, Sun, Zhang, and Overwijk}]{yu2022cocodr}
Yue Yu, Chenyan Xiong, Si~Sun, Chao Zhang, and Arnold Overwijk. 2022.
\newblock Coco-dr: Combating distribution shifts in zero-shot dense retrieval
  with contrastive and distributionally robust learning.
\newblock In \emph{Proceedings of the 2022 Conference on Empirical Methods in
  Natural Language Processing}.

\bibitem[{Zhang et~al.(2021)Zhang, Ma, Shi, and Lin}]{mrtydi}
Xinyu Zhang, Xueguang Ma, Peng Shi, and Jimmy Lin. 2021.
\newblock {Mr. TyDi}: A multi-lingual benchmark for dense retrieval.
\newblock \emph{arXiv:2108.08787}.

\end{thebibliography}
\bibliographystyle{acl_natbib}

\onecolumn
\appendix

\section{Appendix}
\label{sec:appendix}
\subsection{Instructions}
\label{sec:app-inst}
\subsubsection{Web Search}

\begin{tcolorbox}
Please write a passage to answer the question

Question: [QUESTION]

Passage:
\end{tcolorbox}

\subsubsection{SciFact}

\begin{tcolorbox}
Please write a scientific paper passage to support/refute the claim

Claim: [Claim]

Passage:
\end{tcolorbox}

\subsubsection{Arguana}

\begin{tcolorbox}
Please write a counter argument for the passage

Passage: [PASSAGE]

Counter Argument:
\end{tcolorbox}

\subsubsection{TREC-COVID}

\begin{tcolorbox}
Please write a scientific paper passage to answer the question

Question: [QUESTION]

Passage:
\end{tcolorbox}

\subsubsection{FiQA}

\begin{tcolorbox}
Please write a financial article passage to answer the question

Question: [QUESTION]

Passage:
\end{tcolorbox}

\subsubsection{DBPedia-Entity}
\begin{tcolorbox}
Please write a passage to answer the question.

Question: [QUESTION]

Passage:
\end{tcolorbox}

\subsubsection{TREC-NEWS}
\begin{tcolorbox}
Please write a news passage about the topic.

Topic: [TOPIC]

Passage:
\end{tcolorbox}

\subsubsection{Mr.TyDi}
\begin{tcolorbox}
Please write a passage in Swahili/Korean/Japanese/Bengali to answer the question in detail.

Question: [QUESTION]

Passage:
\end{tcolorbox}

\end{document}